\documentclass[useAMS,usenatbib]{mn2e}
\usepackage{graphicx}
\title[ETHOS~1]{
ETHOS~1: A high latitude planetary nebula with jets forged by a post common envelope binary central star\thanks{Based on observations made with the Flemish Mercator and Isaac Newton Telescopes of the Observatorio del Roque de Los Muchachos and the Very Large Telescope at Paranal Observatory under programs 083.D-0654(A) and 085.D-0629(A).}}
\author[B. Miszalski et al.]{B. Miszalski$^{1}$\thanks{E-mail: b.miszalski@herts.ac.uk}, R. L. M. Corradi$^{2,3}$, H. M. J. Boffin$^{4}$, D. Jones$^{5}$, L. Sabin$^{6}$,\newauthor M. Santander-Garc\'ia$^{7,2,3}$, P. Rodr\'iguez-Gil$^{7,2,3}$ and M. M. Rubio-D\'iez$^{7,8}$\\
$^{1}$Centre for Astrophysics Research, STRI, University of Hertfordshire, College Lane Campus, Hatfield AL10 9AB, UK\\
$^{2}$Instituto de Astrof\'isica de Canarias, E-38200 La Laguna, Tenrife, Spain\\
$^{3}$Departmento de Astrof\'isica, Universidad de La Laguna, E-38205 La Laguna, Tenerife, Spain\\
$^{4}$European Southern Observatory, Alonso de Cordova 3107, Casilla 19001, Santiago, Chile\\
$^{5}$Jodrell Bank Centre for Astrophysics, School of Physics and Astronomy, University of Manchester, M13 9PL, UK\\
$^{6}$Instituto de Astronom\'ia, Universidad Nacional Aut\'onoma de M\'exico, Apdo. Postal 877, 22800 Ensenada, B.C, Mexico\\
$^{7}$Isaac Newton Group of Telescopes, Apart. de Correos 321, 38700 Santa Cruz de la Palma, Spain\\
$^{8}$Centro de Astrobiologia, CSIC-INTA, Ctra de Torrej\'on a Ajalvir km 4, E-28850 Torrej\'on de Ardoz, Spain
}

\begin{document}

\date{Accepted . Received ; in original form }

\maketitle

\begin{abstract}
   We report on the discovery of ETHOS~1 (PN G068.1$+$11.0), the first spectroscopically confirmed planetary nebula (PN) from a survey of the SuperCOSMOS Science Archive for high-latitude PNe. ETHOS~1 stands out as one of the few PNe to have both polar outflows (jets) travelling at $120\pm10$ km/s and a close binary central star. The lightcurve observed with the Mercator telescope reveals an orbital period of 0.535 days and an extremely large amplitude (0.816 mag) due to irradiation of the companion by a very hot pre-white dwarf. ETHOS~1 further strengthens the long suspected link between binary central stars of planetary nebulae (CSPN) and jets. INT IDS and VLT FORS spectroscopy of the CSPN reveals weak N III, C III and C IV emission lines seen in other close binary CSPN and suggests many CSPN with these weak emission lines are misclassified close binaries. We present VLT FORS imaging and Manchester Echelle Spectrometer long slit observations from which a kinematic model of the nebula is built. An unusual combination of bipolar outflows and a spherical nebula conspire to produce an $X$-shaped appearance. The kinematic age of the jets ($1750\pm250$ yrs/kpc) are found to be older than the inner nebula ($900\pm100$ yrs/kpc) consistent with previous studies of similar PNe. Emission line ratios of the jets are found to be consistent with reverse-shock models for fast low-ionisation emitting regions (FLIERS) in PNe. Further large-scale surveys for close binary CSPN will be required to securely establish whether FLIERS are launched by close binaries.
\end{abstract}

\section{Introduction}
In the protracted debate on the shaping mechanisms of PNe (Balick \& Frank 2002) appropriate solutions are sought to explain both the dominant nebula morphology (e.g. spherical, elliptical, bipolar; Balick 1987) and the accompanying collimated outflows or `jets'. Single star evolution readily explains spherical PNe and potentially elliptical PNe if there is interaction with the interstellar medium (ISM; Villaver et al. 2003; Wareing et al. 2007). Highly axisymmetric or bipolar nebulae (Corradi \& Schwarz 1995) have been modelled with some success including the generalised interacting stellar wind model (GISW; Kwok, Purton \& Fitzgerald 1978; Kwok 2000) and a combination of stellar rotation and magnetic fields (e.g. Garc\'ia-Segura 1997). There are however fundamental limitations in both of these models. The GISW model depends upon an \emph{assumed} density contrast to produce bipolar PNe (e.g. a dusty torus) and the latter cannot operate without an additional supply of angular momentum (e.g. a binary companion; Soker 2006; Nordhaus, Blackman \& Frank 2007). While there are multiple advantages to a binary explanation over existing theories, we are only beginning to understand the prevalence and impact of binarity in PNe (De Marco 2009; Miszalski et al. 2009a, 2009b, 2010).

Decisive observational evidence is required to advance the shaping debate which is imbalanced towards theoretical models. PNe with dusty disks seem to support the GISW model though there appears to be a strong dependence on a binary companion for their formation (Chesneau 2010). 
Similarly, magnetic fields are difficult to observe directly in PNe (Jordan et al. 2005; Sabin et al. 2007), yet they appear to be more influential as part of a binary-driven dynamos (Nordhaus \& Blackman 2006; Nordhaus et al. 2007). Evidence for close binary central stars of PNe (CSPN) that passed through a common-envelope (CE) phase is now firmly established and gaining considerable momentum with at least 40 now known (Miszalski et al. 2009a, 2010; De Marco et al. 2008). These make up at least $17\pm5$\% of all CSPN (Miszalski et al. 2009a; Bond 2000).

With a larger sample of 30 post-CE nebulae Miszalski et al. (2009b) found at least 30\%, perhaps 60--70\%, of post-CE binaries had bipolar nebulae suggesting the CE phase preferentially shapes bipolar nebulae. This is a substantial improvement over the study of Bond \& Livio (1990) whose sample of around a dozen close binaries showed no clear morphological trends. High resolution kinematics of nebulae around post-CE CSPN have shown a tendency for matched orbital and nebulae inclinations that strengthen the connection between the binary and its nebula (e.g. Mitchell et al. 1997; Jones et al. 2010). Further progress in this aspect will require a statistically significant sample of post-CE nebulae to be identified and studied in detail to reveal trends bestowed upon the nebula during the CE phase (Miszalski et al. 2010). Miszalski et al. (2009b) made the first steps in this direction by finding a propensity for low-ionisation structures (Gon{\c c}alves et al. 2001) amongst post-CE nebulae either in the orbital plane as a ring (e.g. Sab 41, Miszalski et al. 2009b; the Necklace, Corradi et al. 2010) and in the polar direction as jets (e.g. A~63, Mitchell et al. 2007). 

Jets occur in a wide variety of astrophysical objects including PNe in which they are the least understood (Livio 1997). The connection between jets and binarity has long been suspected (Soker \& Livio 1994) but never proven given the paucity of known PNe with jets \emph{and} binary CSPN. 
Jets in PNe typically have low outflow velocities ($\sim$100 km/s), but higher velocities of 200--300 km/s are not uncommon, e.g. M 1-16 (Schwarz 1992; Corradi \& Schwarz 1993), NGC 6337 (Corradi et al. 2000) and NGC 2392 (Gieseking, Becker \& Solf 1985), and can even reach 630 km/s in the case of MyCn 18 (O'Connor et al. 2000). There is no strong evidence for highly collimated outflows (e.g. Herbig-Haro-like outflows) in genuine PNe and Frew \& Parker (2010) suggest the few objects with the most collimated outflows are instead related to B[e] stars. The most striking example is Hen 2-90 (Sahai \& Nyman 2000) which was later reclassified as a B[e] star (Kraus et al. 2005; Frew \& Parker 2010).

Instead, jets in PNe manifest as point-symmetric `corkscrew'-like outflows (L\'opez et al. 1993; V\'azquez et al. 2008) or as pairs of opposing knots often called fast low-ionisation emitting regions (FLIERS; see e.g. Balick et al. 1987, 1993; Dopita 1997) with notable examples being Hb 4 (L\'opez et al. 1997; Harrington \& Borkowski 2000; Miszalski et al. 2009b), NGC 6337 (Corradi et al. 2000; Garc\'ia-D\'iaz et al. 2009), A~63 (Mitchell et al. 2007), IC 4634 (Guerrero et al. 2008), IC 4673 (Kovacevic et al. 2010), Sab 41 (Miszalski et al. 2009b) and the Necklace (Corradi et al. 2010). Further examples are listed by Gon{\c c}alves et al. (2001) and Harrington \& Borkowski (2000).

Point-symmetric jets have received the most attention in the literature with models consisting of an episodic precessing jet powered by an accretion disk in a wide binary with $P_\mathrm{orb}$$\sim$a few years (e.g. Raga, Cant\'o \& Biro 1993; Cliffe et al. 1995; Haro-Corzo et al. 2009; Raga et al. 2009). 
Precession helps smear out or widen the jet and may also explain the often `bent' position of jets with respect to the presumed major axis (e.g. Hb 4 and Sab 41). Some jets appear in two pairs (e.g. NGC 6337) and these may have arose from multiple ejections (e.g. Nordhaus \& Blackman 2006). A similar scenario involves a \emph{short-lived} `wobbling' accretion disk (Livio \& Pringle 1996, 1997; Icke 2003) and models with magnetic fields have also successfully been applied to the problem (Garc\'ia-Segura 1997). Dynamical lifetimes of jets suggest they are ejected well before the main nebula (Mitchell et al. 2007; Corradi et al. 2010), i.e. before the CE is ejected if a close binary is present, and this is consistent with the short-lived nature of the modelled accretion disks.

The most constructive way to establish the link between jets and binaries is to look for binary CSPN in a large sample of PNe with jets. This is one of the major aims behind our ongoing program to look for close binary CSPN via the photometric monitoring technique (Miszalski et al. 2010). We are also targeting bipolar PNe and through non-detections we will also be able to discern whether larger orbital periods produce point-symmetric jets rather than FLIERS. The program has been very successful so far with at least five close binary CSPN found using the 1.2-m Flemish Mercator telescope. These include the Necklace and its Supernova 1987A-like ring (Corradi et al. 2010), the Type Ia supernova candidate double-degenerate nucleus of Hen 2-428 (Santander-Garc\'ia et al. 2010), and the subject of this paper, ETHOS~1. 

This paper is structured as follows. Sec. \ref{sec:discovery} introduces the new survey responsible for the discovery of ETHOS~1 and the spectroscopic confirmation of its PN status. Spectroscopic and photometric evidence for binarity are presented in Sec. \ref{sec:binary}. The nebula is examined in Sec. \ref{sec:neb} with narrow-band images, high-resolution spectroscopy. Sec. \ref{sec:diag} examines diagnostic emission line ratios of the jets and we conclude in Sec. \ref{sec:conclusion}. 

\section{Discovery and PN Confirmation}
\label{sec:discovery}
\subsection{The Extremely Turquoise Halo Object Survey (ETHOS)}
Beyond the confines of Galactic Plane H$\alpha$ surveys (Parker et al. 2005; Drew et al. 2005) there are very few PNe currently known as less than 9\% of $\sim$2800 Galactic PNe are located at latitudes $|b|\ga10^\circ$. Most of these were found from visual inspection of photographic broadband sky survey plates (e.g. Kohoutek 1963; Abell 1966; Longmore 1977; Weinberger 1977; Longmore \& Tritton 1980; Hartl \& Tritton 1985). With the digitisation of the same material it is now possible to extend this work to lower surface brightnesses and to systematically search larger areas more efficiently (Jacoby et al. 2010). The MASH-II project (Miszalski et al. 2008) demonstrated that visual searches for PNe are largely insensitive to compact or star-like PNe. Several compact PNe were found from catalogue queries of the SuperCOSMOS H$\alpha$ Survey (Parker et al. 2005) including the MASH-II PN MPA~1327-6031 that lies just near the outer lobe of the bipolar MASH PN PHR~1327-6032 (Parker et al. 2006). The number of distant (compact) Halo PNe is therefore essentially unknown. Further discoveries are essential to determine whether Halo PNe can only be formed via close binary evolution (Bond 2000) as there is currently only a small amount of supporting evidence (e.g. Pe\~na \& Ruiz 1998; Tovmassian et al. 2010) 

\begin{figure}
   \begin{center}
      \includegraphics[scale=0.55]{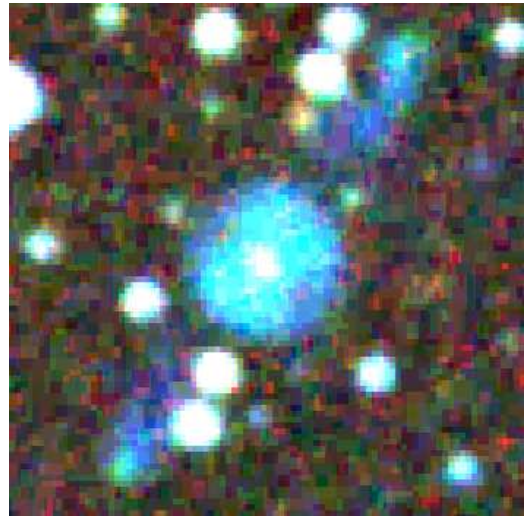}
   \end{center}
   \caption{Discovery image of ETHOS~1 made from $I_N$ (red), $R_F$ (green) and $B_J$ (blue) POSS-II SuperCOSMOS Sky Survey images (Reid et al. 1991; Hambly et al. 2001a). The turquoise hue is typical of PNe in this particular colour-composite image combination. North is to top and East to left in this 1 $\times$ 1 arcmin$^2$ image.}
   \label{fig:discovery}
\end{figure}

The \emph{Extremely Turquoise Halo Object Survey} (ETHOS, Miszalski et al. in prep) is a new survey designed to find high-latitude compact PNe that would ordinarily be missed by visual searches. 
ETHOS is essentially an extension to broadband colours of the MASH-II catalogue search applied to the whole sky (excluding the H$\alpha$ surveyed regions). Currently four candidate PNe have been spectroscopically confirmed out of a few dozen candidates identified solely from queries of the SuperCOSMOS Science Archive (SSA, Hambly et al. 2004), a convenient \textsc{SQL} interface to the entire database of catalogue data generated during the digitisation of SuperCOSMOS Sky Survey data (Hambly et al. 2001b). 
Candidates are then visualised as for MASH-II, but this time replacing the primary colour-composite H$\alpha$ (red), SR (green) and $B_J$ (blue) image with an $I_N$ (red), $R_F$ (green) and $B_J$ (blue) image (Reid et al. 1991; Hambly et al. 2001a). 
The \emph{IRB} image is particularly effective in distinguishing PNe apart from other objects since they often appear with a strong turquoise hue. Figure \ref{fig:discovery} best shows the turquoise false-colour in the discovery image of ETHOS~1, the first PN candidate identified during the design phase of ETHOS (Miszalski 2009). The object was selected based on the colours of the inner nebula but the presence of faint extensions either side made it stand out amongst the other candidates. Further details of the survey and initial results will be presented elsewhere (Miszalski et al. in prep).

\subsection{Spectroscopic confirmation of ETHOS~1}
ETHOS~1 was observed during Isaac Newton Telescope (INT) service time with the Intermediate Dispersion Spectrograph (IDS) on 10 March 2009. The R300V grating was used at a central wavelength of 5400 \AA\  to provide wavelength coverage from 3300--8000 \AA\ at a dispersion of 1.87 \AA/pixel. The spectrograph slit was placed along the jets at a position angle (PA) of $149^\circ$ and its width of 1.0\arcsec\ gave a resolution of 5 \AA\ (FWHM) at 5400 \AA. One 30 minute exposure was taken at an airmass of 1.51 followed by a 30 s exposure of the spectrophotometric standard star HD 192281 at an airmass of 1.72 with an 8\arcsec\ slit. Standard reductions were performed with \textsc{IRAF} and four different extractions corresponding to different nebula regions were made using the \textsc{IRAF} task \textsc{apall}. The stellar continuum of the CSPN was traced to guide the extractions. Aperture sizes used were 6 pixels for the CSPN (2.4\arcsec), 54 pixels for the inner nebula (21.6\arcsec, which includes the CSPN), 19 pixels for the SE jet (7.6\arcsec) and 23 pixels for the NW jet (9.2\arcsec). The jet extraction widths were measured from the jet tips inward towards the CSPN. The extractions were then flux calibrated in the standard fashion with the aforementioned standard. 

\begin{figure*}
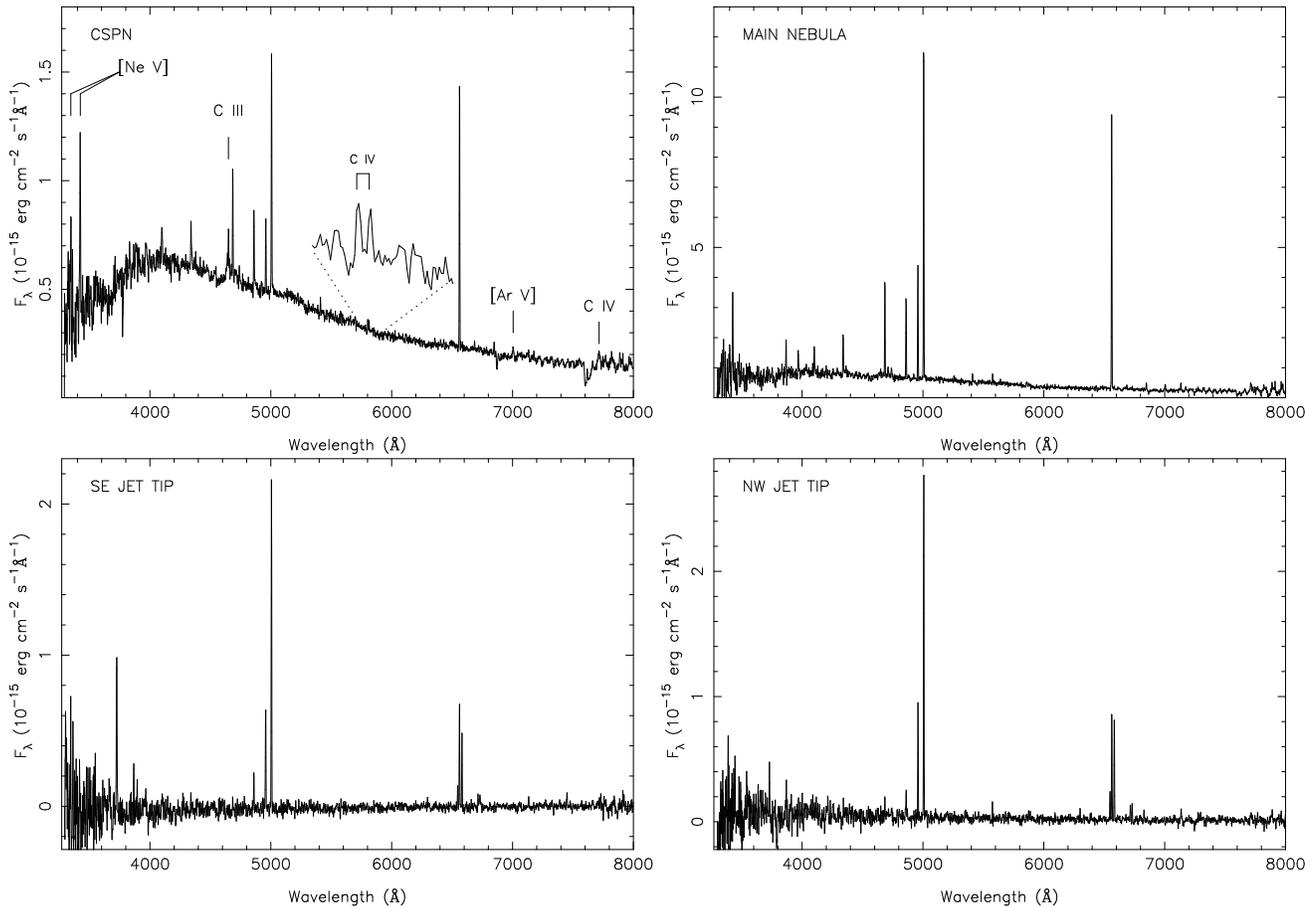

   \begin{center}
      \includegraphics[scale=0.35,angle=270]{IDS_CS.ps}
      \includegraphics[scale=0.35,angle=270]{IDS_MAIN.ps}
      \includegraphics[scale=0.35,angle=270]{IDS_JBTM.ps}
      \includegraphics[scale=0.35,angle=270]{IDS_JTOP.ps}
   \end{center}
   \caption{INT IDS spectroscopy of ETHOS~1. No nebular subtraction was made for the CSPN spectrum.}
   \label{fig:ids}
\end{figure*}

Figure \ref{fig:ids} shows the IDS spectra that confirm the PN nature of ETHOS~1 (PN~G068.1$+$11.0) whose basic properties are given in Tab. \ref{tab:basic}.
We measured $c$, the logarithmic extinction at H$\beta$, to be 0.18 using the Howarth (1983) reddening law and $T_e=17700$ K (Sec. \ref{sec:diag}).
The high ratio of He~II $\lambda$4686/H$\beta\ga1.1$ in the main nebula is typical of very high excitation PNe at high Galactic latitudes in which all the helium is doubly ionised giving a mass-bounded, optically thin nebula (Kaler 1981; Kaler \& Jacoby 1989). 
Similar emission line intensities are also found in the post-CE PNe K 1-2 (Exter et al. 2003) and the Necklace (Corradi et al. 2010), but in our case we do not find any low-ionisation species in the main nebula perhaps suggesting a slightly higher CSPN temperature. The only low-ionisation species are found in the tips of the jets which are caused by shocks (Sec. \ref{sec:diag}). Besides the [Ne V] and [Ar V] emission found close to the hot CSPN, the CSPN spectrum also shows C III $\lambda$4650, C IV $\lambda$5801, $\lambda$5812 and $\lambda$7726 as the signature of an irradiated close companion (Sec. \ref{sec:irrad}).

\begin{table}
   \centering
   \caption{Basic properties of ETHOS~1.}
   \label{tab:basic}
   \begin{tabular}{lr}
      \hline
      PN~G & 068.1$+$11.0\\
      RA (J2000) & $19^\mathrm{h}16^\mathrm{m}31.5^\mathrm{s}$\\
      Declination (J2000) & $+36^\circ09'47.9''$\\
      Diameter (main nebula) & 19.4\arcsec\\
      Jet length (tip-to-tip) & 62.6\arcsec\\
      $c$(H$\beta$) & 0.18 \\
      Heliocentric Radial Velocity & $-20$ km/s\\
      \hline
   \end{tabular}
\end{table}

\section{The Binary Central Star}
\label{sec:binary}
\subsection{An irradiated spectroscopic binary}
\label{sec:irrad}
ETHOS~1 was also observed with the FORS2 instrument (Appenzeller et al. 1998) under the VLT visitor mode program 083.D-0654(A) on 14 July 2009. The objective of this program was to measure the radial velocity curves of binaries discovered by Miszalski et al. (2009a), however inclement weather forced the execution of a backup program in which ETHOS~1 was included. 
The blue optimised E2V detector from FORS1 was used in combination with the 1200g grism to give continuous wavelength coverage from 4058--5556 \AA. The CCD was readout with $2\times2$ binning to give a dispersion of 0.72 \AA/pixel, a spatial resolution of 0.252\arcsec/pixel and a resolution measured from arc lines of 1.57 \AA\ (FWHM) at 4800 \AA.

\begin{figure*}
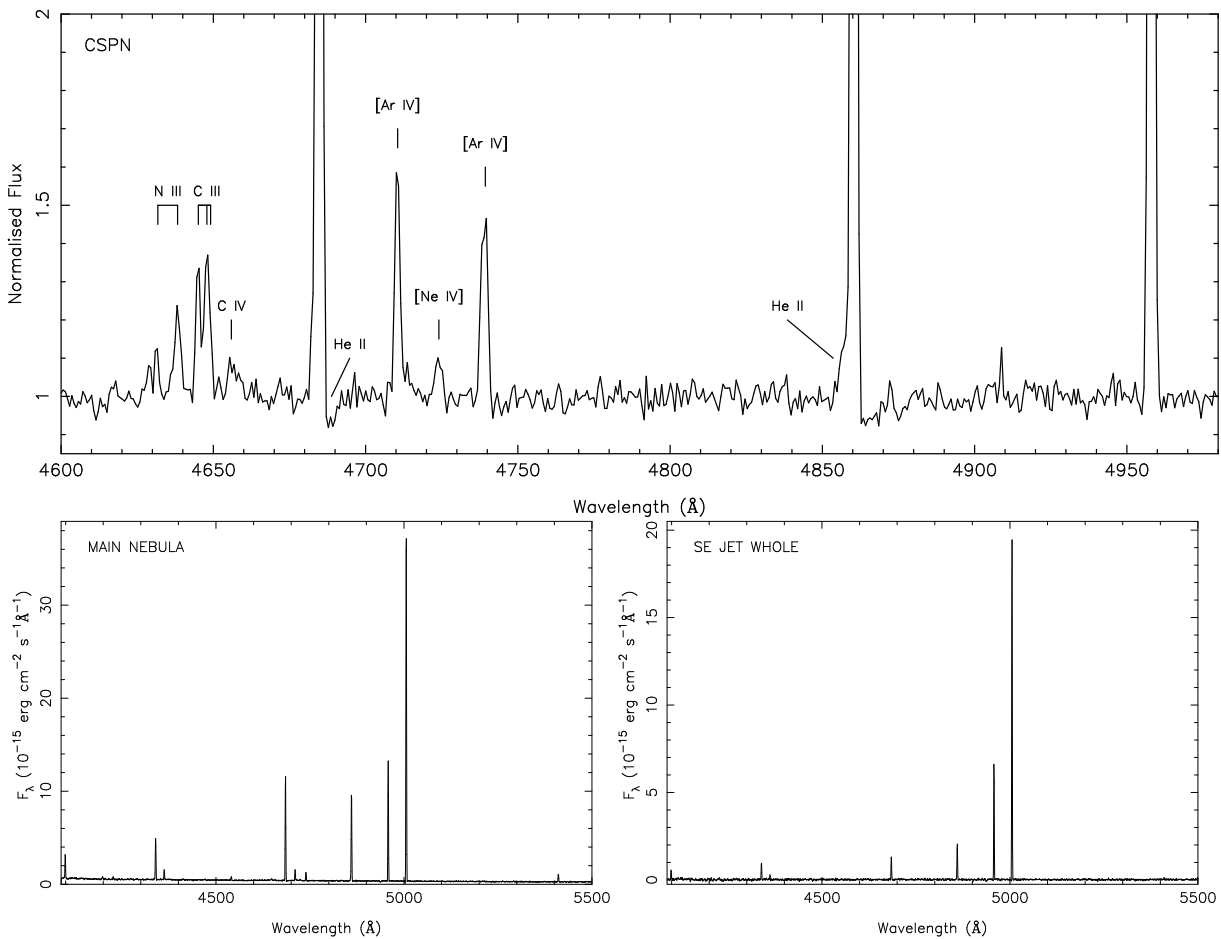

   \begin{center}
      \includegraphics[scale=0.65,angle=270]{ethos_cspn}
      \includegraphics[scale=0.325,angle=270]{vltmain.ps}
      \includegraphics[scale=0.325,angle=270]{vltjet.ps}
   \end{center}
   \caption{VLT FORS spectroscopy of ETHOS~1. No nebular subtraction was made for the rectified CSPN spectrum.}
   \label{fig:vspec}
\end{figure*}

A 40 minute spectrum of ETHOS~1 was obtained at an airmass of 2.04 and the seeing varied between 1.0--1.3\arcsec under thin-thick cloud (relative flux rms up to 0.07) limiting the SNR in the continuum to $\sim$40 near He~II $\lambda$4686. The 0.7\arcsec\  slit was placed at PA $=144^\circ$ to cover the jets but because of the gap between FORS chips not all of the NW jet was observed. The spectroscopic data were processed using the ESO FORS pipeline and the \textsc{iraf} task \textsc{apall} was used to make three extractions using the CSPN continuum as a reference. These were 8 pixels for the CSPN (2.0\arcsec), 84 pixels for the inner nebula (21.2\arcsec, which includes the CSPN) and 70 pixels for the SE jet (17.6\arcsec). The jet extraction widths were measured from the jet tips inward towards the CSPN. Each spectrum was flux calibrated using a 300s exposure of the CSPN of NGC 7293 (Oke 1990) obtained at the end of the same night at an airmass of 1.27 and a slit width of 0.7\arcsec. The final spectra are depicted in Fig. \ref{fig:vspec}.

Features identified in the IDS and FORS CSPN spectra are typical of an irradiated close binary CSPN (Pollacco \& Bell 1993, 1994). The hot primary is revealed by He~II $\lambda$4686, $\lambda$5412 in absorption and the primary is irradiating a main-sequence companion to produce N~III ($\lambda$4634.14, $\lambda$4640.64), C~III ($\lambda$4647.42, $\lambda$4650.25) and C~IV ($\lambda$4658.30, $\lambda$5801.33, 5811.98, $\lambda$7726.20) emission lines. These lines originate from the irradiated zone on the side of the companion facing the primary and therefore serve as a proxy for the motion of the secondary. However, it should be noted that these lines sample the centre of light and not the centre of mass of the secondary (e.g. Exter et al. 2003).

To measure any shifts present we used the [O~III] and H$\beta$ emission lines of our FORS spectrum of the inner nebula as a reference. The systemic heliocentric velocity of the inner nebula was set to be $V_\mathrm{sys}=-20$ km/s from higher resolution spectroscopy (Sec. \ref{sec:morph}). The N~III $\lambda$4640.64, C~III $\lambda$4647.42, $\lambda$4650.25, and C~IV $\lambda$4658.30 emission lines gave an average $V_\mathrm{sec}=-122.0\pm4.6$ km/s. Estimating $V_\mathrm{pri}$ is much more difficult given the prominent nebular He~II emission but the He~II $\lambda$4686 absorption line is present (albeit at low S/N, Fig. \ref{fig:vspec}) and redshifted compared to the nebula velocity. The exact line profile centre is difficult to establish given the nebular emission and the chance that the line core may be partially filled in by irradiated emission. The large uncertainty means we can only estimate $V_\mathrm{pri}\sim105\pm80$ km/s. \emph{Our single epoch spectrum therefore confirms the presence of a spectroscopic binary with both components well separated from each other and the nebula}. Given the large uncertainty in $V_\mathrm{pri}$ and the lack of a well-sampled radial velocity curve it is premature to estimate the masses of each component at this stage. 

An important corollary is that the presence of the N III, C III and C IV emission lines alone would ordinarily be sufficient grounds for a weak emission line star or \emph{wels} classification (Tylenda et al. 1993; Marcolino \& Ara\'ujo 2003). The \emph{wels} are a heterogeneous class of CSPN that are poorly defined (Fogel, De Marco \& Jacoby 2003) and do not fit well-established classification schemes (M\'endez 1991; Crowther et al. 1998; Acker \& Neiner 2003). According to Tylenda et al. (1993), \emph{wels} are identified by (i) C IV $\lambda$5801, 5812 weaker and narrower than in [WC] CSPN, (ii) very weak or undetectable C III $\lambda$5696, and (iii) the emission complex at $\lambda$ 4650. However, this is exactly what is observed in many close binary CSPN with the best examples being Abell 46 (Pollacco \& Bell 1994), ETHOS~1 (Fig. \ref{fig:vspec}) and the Necklace (Corradi et al. 2010). \emph{Many \emph{wels} are therefore likely to be misclassified close binaries (Miszalski 2009)}. Further examples will be presented elsewhere and we expect further high resolution spectroscopic observations of so-called \emph{wels} will identify further close binaries.

\subsection{Lightcurve}
\begin{figure*}
   \begin{center}
      \includegraphics[angle=270,scale=0.55]{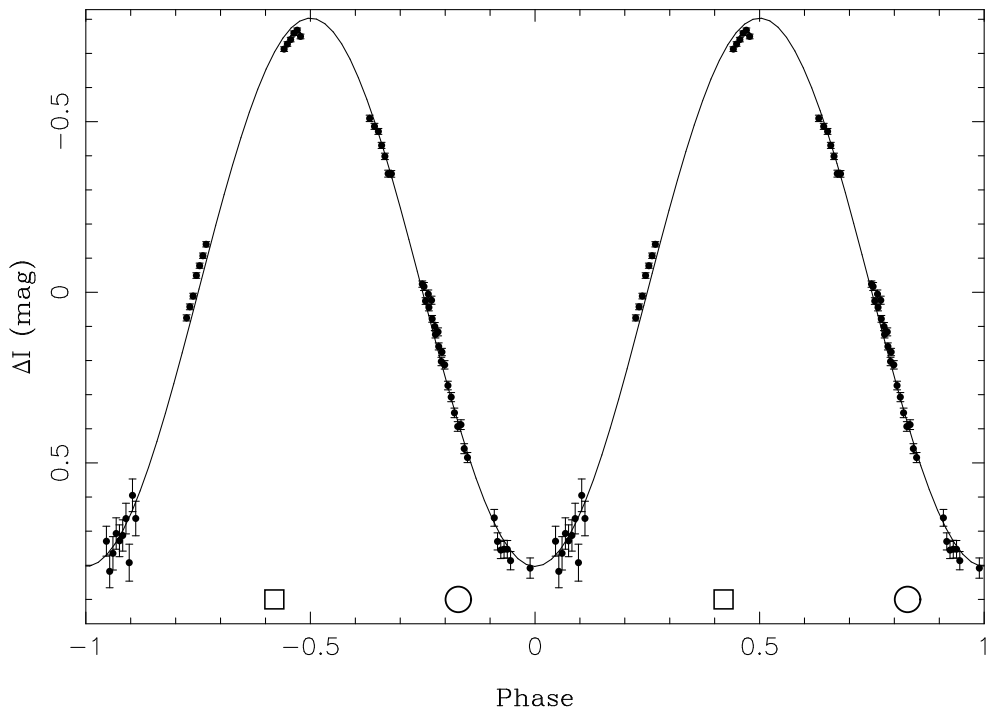}
   \end{center}
   \caption{Mercator lightcurve of ETHOS~1 phased with $P=0.53512$ days. The curve is a sinusoidal fit and phase coverage of IDS (squares) and VLT (circles) spectra are shown.}
   \label{fig:lc}
\end{figure*}

Time series photometry of ETHOS~1 was obtained with the MEROPE camera (Davignon et al. 2004) on the 1.2-m semi-robotic Flemish Mercator Telescope (Raskin et al. 2004) from 24 August 2009 to 4 September 2009. Figure \ref{fig:lc} shows the lightcurve phased with our ephemeris: HJD (min. I) = 2455076.0312$\pm$0.0007 + (0.53512$\pm$0.00019) E. The IDS and FORS spectra taken earlier during the same year covered phases 0.40--0.44 and 0.80--0.85, respectively. As for the Necklace, the observed periodic variability is clear evidence of binarity. A sine fit to the lightcurve gives a very high amplitude of 0.816 mag that indicates a very hot primary irradiating a cool main-sequence companion (see e.g. Fig. 1 of De Marco et al. 2008). Such a large amplitude places ETHOS~1 amongst a small group of post-CE binaries showing extreme irradiation effects with only Sab 41 having a larger amplitude ($I=0.849$ mag, Miszalski et al. 2009a). The next largest amplitudes are found in the Necklace ($I=0.747$ mag, Corradi et al. 2010) and K 1-2 ($V=0.680$ mag, Exter al. 2003). 
There are no other post-CE binaries without visible nebulae that come close to these extreme amplitudes simply because they have evolved significantly further along the white dwarf cooling track (Aungwerojwit et al. 2007; Shimansky et al. 2009). 

\section{The Nebula}
\label{sec:neb}
\subsection{Morphology and Kinematics}
\label{sec:morph}
\begin{figure*}
   \begin{center}
      \includegraphics[scale=0.70]{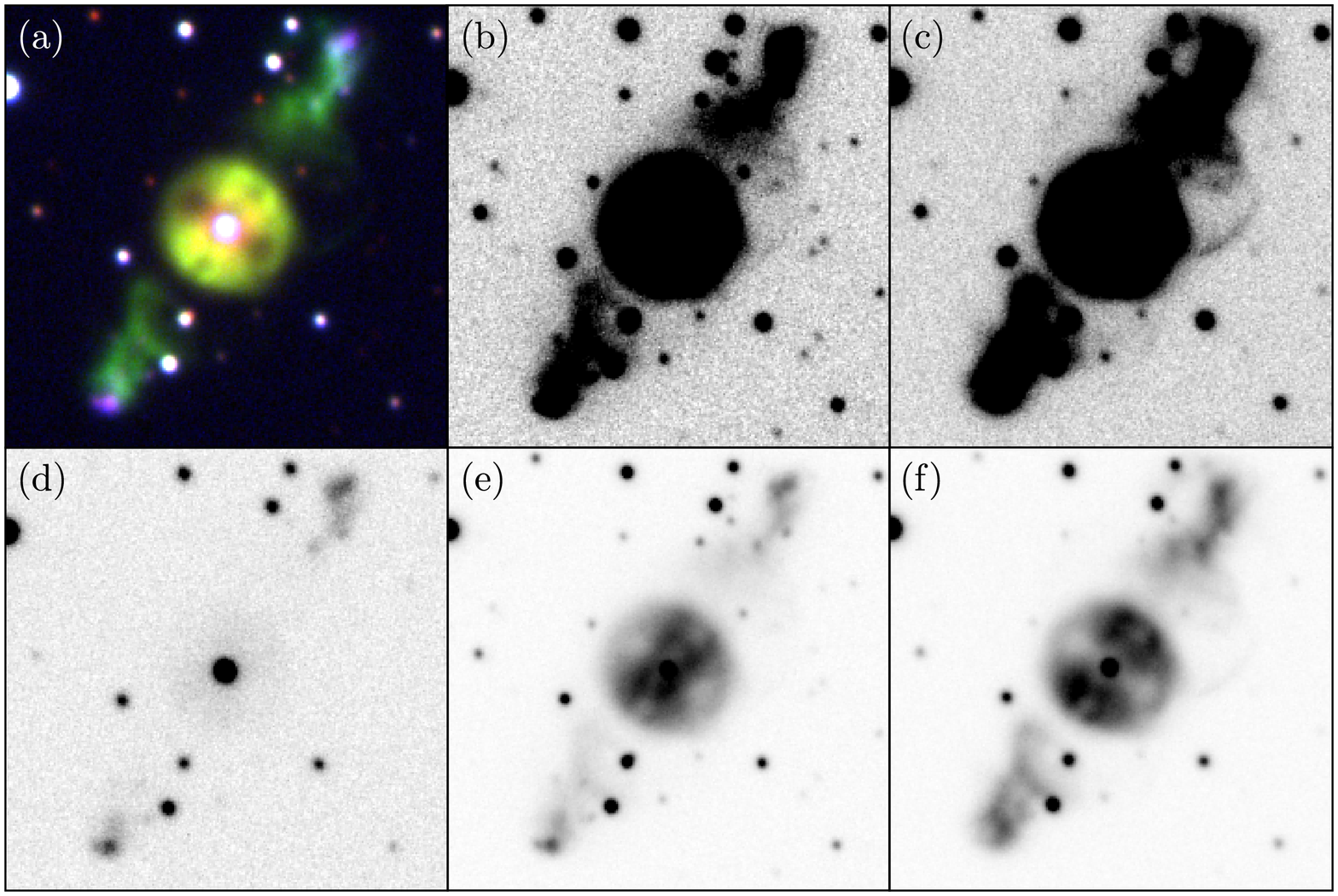}
   \end{center}
   \caption{A montage of VLT FORS ETHOS~1 images. (a) Colour-composite of H$\alpha$+[N~II] (red), [O~III] (green) and [O II] (blue); (b) and (e) H$\alpha$+[N~II]; (c) and (f) [O~III]; (d) [O II]; Each image measures $1\times1$ arcmin$^2$ with North up and East to the left.}
   \label{fig:montage}
\end{figure*}

Narrow-band images of 300 s duration each were obtained during the VLT visitor mode program 085.D-0629(A) on 19 June 2010 with the FORS H\_Alpha$+$83, OIII$+$50 and OII$+$44 filters. The central wavelengths and full widths at half maximum (FWHM) of each filter are 656.3/6.1 nm, 500.1/5.7 nm and 377.6/6.5 nm, respectively. Note the H\_Alpha$+$83 filter includes the [N~II] $\lambda$6548, $\lambda$6583 emission lines. Figure \ref{fig:montage} presents a montage of the images which show three components to ETHOS~1: (i) The main nebula 19.4\arcsec\ across measured at 10\% of the peak H$\alpha$ intensity, (ii) the jets measuring 62.6\arcsec\ tip-to-tip also at 10\% of the peak H$\alpha$ intensity, and (iii) a faint bipolar outflow best visible in [O~III] (Fig. \ref{fig:montage}c). 

The absence of an inner ring of low-ionisation structures (LIS) sets ETHOS~1 apart from the other nebulae surrounding high-amplitude close binaries including K 1-2 (Exter et al. 2003), Sab 41 (Miszalski et al. 2009b) and the Necklace (Corradi et al. 2010). As the UV radiation field is comparably high in the other objects this may indicate the ring never existed rather than recent ablation of an LIS ring.  There are in fact no low-ionisation species detected in ETHOS~1 except for in the shocked jet tips described in Sec \ref{sec:diag}. A faint detection in the [O II] image is due to the nebula continuum since there is no [N~II], [S~II] or [O II] emission recorded by our INT spectrum of the main nebula (Fig. \ref{fig:ids}). 

Spatially resolved long slit echelle spectra of ETHOS~1 were made using the second Manchester Echelle Spectrometer on the 2.1-m San Pedro M\'artir Telescope (Meaburn et al. 2003). The primary spectral mode was used with narrowband filters to isolate the [O~III] $\lambda$5007, H$\alpha$ and [N~II] $\lambda$6548, 6583 emission lines. Observations were performed on 23 September 2010 with a 2\arcmin\ $\times$ 2\arcsec\ slit and the Thomson 2K$\times$2K CCD which gave a spatial scale of 0.38\arcsec/pixel and a resolving power of $R\sim30000$ (2.82 km/s per pixel). A series of 1800 s exposures were taken with the slit oriented approximately along the minor (PA=60$^\circ$) and major (PA=147$^\circ$) axes. 

Figure \ref{fig:echelle} presents the emission line position-velocity (PV) diagrams observed. Alongside these are the same PV diagrams extracted from a \textsc{shape} (Steffen \& L\'opez 2006) best-fitting spatio-kinematic model of ETHOS~1 built following Jones et al. (2010). Figure \ref{fig:shape} shows the model that was inspired by the faint Hb 12-like bipolar extensions (Vaytet et al. 2009) that are visible in Fig. \ref{fig:montage}c. The basic components are a spherical outflow and nested thin-walled bipolar outflows. The inner bipolar outflow helps reproduce the $X$-shape and the outer (possibly related) Hb 12-like bipolar outflow protrudes from the central sphere. Our approach was not tailored to each emission line observed, but instead aimed to create a good overall approximation to the features seen in Figs. \ref{fig:montage} and \ref{fig:echelle}. The jets were not incorporated in our model but they appear to be well-aligned to the major axis unlike other PNe (e.g. Hb4). The faint bipolar extensions (Fig. \ref{fig:montage}c) were included in our model but their low surface brightness precluded their detection on the 2.1-m telescope at this resolution. We note that no attempt has been made to replicate the relative surface brightnesses of each component in the model and deeper observations should be able to detect these fainter features.

\begin{figure*}
   \begin{center}
      \includegraphics[scale=0.8,angle=270]{Dave_echelle.ps}
   \end{center}
   \caption{Model and observed position-velocity diagrams for the major (PA=147$^\circ$) and minor (PA=60$^\circ$) axes of ETHOS~1. The velocity scale is centred on the model-derived heliocentric radial velocity of $-20$ km/s. The jets are not incorporated into the models and our observations are not sensitive to the faint bipolar outflow (major axis model). The orientation of the major and minor axes are North is up and East is up, respectively.}
   \label{fig:echelle}
\end{figure*}

\begin{figure}
   \begin{center}
      \includegraphics[scale=0.75]{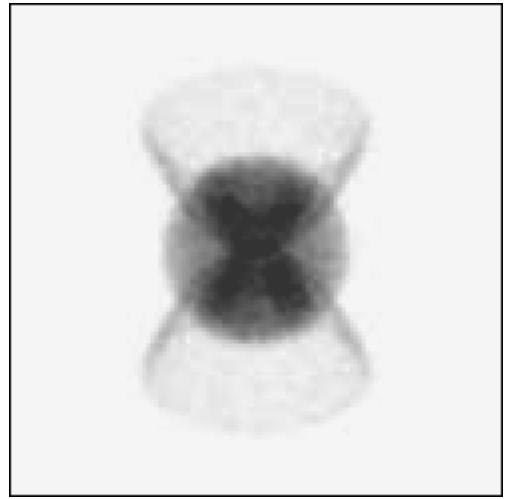}
   \end{center}
   \caption{A \textsc{shape} spatio-kinematic model of ETHOS~1.}
   \label{fig:shape}
\end{figure}

Along the major axis the PV diagrams of the inner nebula display a figure-of-eight profile best seen in H$\alpha$ that is typical of a bipolar outflow with a `pinched' waist (e.g. Icke, Preston \& Balick 1989). This bipolar profile is also apparent in [O~III], but is far more irregular and lacks emission from the central region. The southern component splits into two velocity components that are consistent with the front and back walls of a hollow bipolar shell. A velocity ellipse is also seen in the [O~III] profile surrounding the figure-of-eight profile. This is consistent with a spherical shell of material and the minor axis PV diagrams support this interpretation. Along the minor axis the outer ellipse represents the spherical shell, while the inner knot at negative velocity represents the pinched bipolar shell (e.g. Vaytet et al. 2009). There is a small amount of asymmetry in the minor axis which is partly due to imperfect alignment of the slit along the axis and partly intrinsic.

No emission is detected from the central region in [N~II] consistent with our other observations. There is however an elongated excess of H$\alpha$ emission in the inner nebula aligned along the major axis that is not matched in [O~III] (e.g. Fig. \ref{fig:montage}a). While this could be explained by ionisation stratification (e.g. the same feature may be visible in O~V or O~VI), it may also be related to similar but `tilted' inner nebulae in PN~G126.6$+$01.3 (Mampaso et al. 2006), M 2-19 (Miszalski et al. 2009b) and A 41 (Miszalski et al. 2009b; Jones et al. 2010). We suggest these nebulae could be the remnants of a precessing jet (Haro-Corzo et al. 2009; Raga et al. 2009) or nozzle (Balick 2000). The uncanny resemblance between PN~G126.6$+$01.3 and the models of Haro-Corzo et al. (2009) supports this hypothesis. 

The symmetry axis of our best-fitting \textsc{shape} model has an inclination of $60\pm5$ degrees to the line of sight. This inclination is consistent with the non-eclipsing lightcurve, the high degree of mirror symmetry evident in the main nebula about the minor axis and the large visibility fraction of the jets.
A heliocentric radial velocity of $-20\pm5$ km/s was determined from our model for the nebula. 
The expansion of the model nebula follows a Hubble-type flow, where all velocities are proportional to the radial distance from the nebular centre, equivalent to an expansion velocity of 55 km/s at the edge of the spherical component. This is above average for a PN but not unexpected for a bipolar outflow (Corradi \& Schwarz 1995) with He~II $\lambda$6560 present (Richer et al. 2008). Figure \ref{fig:vspec} shows He~II $\lambda$4859 and we confirm the detection of He~II $\lambda$6560 in our MES spectroscopy.

Although the distance to ETHOS~1 is unknown, the spatio-kinematic model was used to determine the kinematical age per kpc of the inner nebula as $900\pm100$ years kpc$^{-1}$. Similarly, assuming the jets have the same inclination as the internal bipolar structure, we find their kinematical age to be $1750\pm250$ years kpc$^{-1}$ at their tips, almost double that of the central region! 
The velocities of the jet tips reach $-55\pm5$ km/s (SE jet) and $65\pm5$ km/s (NW jet). Their deprojected velocities at our assumed inclination are $-110\pm10$ km/s and $130\pm10$ km/s, respectively.

\begin{table}
   \centering
   \caption{Distance dependent kinematic ages of nebulae and jets of PNe with close binary nuclei.}
   \label{tab:ages}
   \begin{tabular}{crrc}
      \hline 
      Name & $t_\mathrm{nebula}$ & $t_\mathrm{jets}$ & Reference\\
           &     (yrs/kpc)                  &   (yrs/kpc)  &   \\
      \hline 
      A~63 & $3500\pm200$     & $5200\pm1200$  & Mitchell et al. (2007)\\
      Necklace &  $1100\pm100$  & $2350\pm450$ & Corradi et al. (2010)\\
      ETHOS~1 & $900\pm100$ & $1750\pm250$ & this work\\
      \hline
   \end{tabular}
\end{table}

\subsection{Emission line diagnostics}
\label{sec:diag}
Table \ref{tab:lines} gives our measured emission line intensities for our extracted IDS and FORS spectra. Dereddened intensities were calculated using the Howarth (1983) reddening law and $c=0.18$. As Exter et al. (2003) point out an abundance analysis for such high excitation nebulae requires photoionisation modelling to properly measure the chemical abundances. Since our spectral information is rather limited and lacks reliable UV and far-blue coverage we postpone this work for now. Abundances of the jets are likely heavily distorted by their shocked nature, but we can investigate their emission line ratios for diagnostic purposes. 

The \textsc{nebular} package in \textsc{iraf} (Shaw \& Dufour 1995) was used to determine electron temperatures and densities in the main nebula and SE jet. Assuming $n_e=1000$ cm$^{-3}$, the main nebula has a very hot $T_e=17700\pm500$ K. This is comparable to K 1-2 with $T_e=17500\pm1000$ K, but is larger than $T_e=14800^{+530}_{-460}$ K measured in the Necklace (Corradi et al. 2010). The [Ar IV] $\lambda$4711/4740 density diagnostic from our VLT spectrum lies near the lower density limit so we find $n_e=850\pm1000$ cm$^{-3}$. We also find the jets to be cooler $T_e=12900\pm1000$ K and to have a similarly low density by using the [S~II] $\lambda$6731/6717 diagnostic to provide an upper limit of $n_e\la1000$ cm$^{-3}$.

\begin{table*}
   \centering
   \caption{Measured ($F_\lambda$) and dereddened ($I_\lambda$) nebula emission line intensities using $c=0.18$.}
   \label{tab:lines}
   \begin{tabular}{lrrrrrrrrrr}
      \hline
      %SE = JBTM; NW = JTOP
      Component: & \multicolumn{2}{c}{Main Nebula} & \multicolumn{2}{c}{SE Jet Tip} & \multicolumn{2}{c}{NW Jet Tip} & \multicolumn{2}{c}{Main Nebula} & \multicolumn{2}{c}{SE Jet Whole} \\
      Instrument: & \multicolumn{2}{c}{IDS} & \multicolumn{2}{c}{IDS} & \multicolumn{2}{c}{IDS} & \multicolumn{2}{c}{FORS} & \multicolumn{2}{c}{FORS} \\
       Identification  &  $F_\lambda$ & $I_\lambda$ &  $F_\lambda$ & $I_\lambda$ &  $F_\lambda$ & $I_\lambda$ &  $F_\lambda$ & $I_\lambda$ &  $F_\lambda$ & $I_\lambda$  \\
         \hline
        {}[Ne V] 3425$^\dag$ & 122 & 140 & -& -& -& - & - & - & - & -  \\
        {}[O II] 3727$^\dag$ &- &- & 642 & 713 & 133& 148 & - & - & - & -  \\
        {}[Ne III] 3869$^\dag$ & 38 & 42 & 133 & 147& 67& 73 & - & - & - & - \\
        {}[Ne III]+HeI/II 3968$^\dag$ & 38 & 41 & - & - & - & - & - & - & - & -  \\
        {}H$\delta$ 4101$^\dag$ & 40 & 44 & - & - & - & - & 25.8& 27.8& 28.8 & 31.0  \\
        {}He~II 4200 &- & - & - & - & - & -  & 4.3 & 4.6 & - & -  \\
        {}C II 4267 &- & - & - & - & - & -  & 3.2 & 3.4 & - & -  \\
        {}H$\gamma$ 4340 & 51 & 54 & - & - & - & -  & 45.7 & 48.1 & 45.7 & 48.2  \\
        {}[O~III] 4363 & $*$ & $*$& - & - & - & - &9.7  & 10.3  & 11.1: & 11.7: \\
        {}He I 4471  &- & - & - & - & - & -  & -   & -   & 6.4: & 6.6:  \\
        {}He~II 4541 &- & - & - & - & - & -  & 3.9 & 4.0 & -     & -  \\
        {}He~II 4686 & 115 & 117 & 20: & 21: & 50: & 51: & 111.8 & 113.8 & 61.1 & 62.2\\
        {}[Ar IV] 4711 &$*$ & $*$ & - & - & - & -  & 11.8 & 12.0 & - & - \\
        {}[Ne IV] 4725 &- & - & - & - & - & -  &  2.2 & 2.2 &- & - \\
        {}[Ar IV] 4740 &$*$ & $*$ & - & - & - & -  & 9.0 & 9.1 & - & - \\
        {}H$\beta$ 4861 & 100 & 100 & 100 & 100 & 100 & 100 & 100.0 & 100.0 & 100.0 & 100.0\\
        {}[O~III] 4959 & 128& 127&  317 & 313&317 & 313 & 132.5 & 131.2 & 299.5 & 296.6 \\
        {}[O~III] 5007 & 404& 398& 975 & 960& 960& 946 & 393.1 & 387.3 & 891.5 & 878.3 \\
        {}He~II 5412 & 10: & 10: & - & - & - & -  & 8.9 & 8.4 & 4.2: & 4.0: \\
        {}[O I] 6300 &- &- & $*$ & $*$& 33: & 30: & - & - & - & - \\
        {}[Ar V] 6434 &5: & 1: & - & - & - & -  & - & - & - & - \\
        {}[N~II] 6548 &- &- & 57 & 50 & 79 & 69 & - & - & - & - \\
        {}H$\alpha$ 6563 & 317& 278& 304 & 266 & 293 & 257 & - & - & - & - \\
        {}[N~II] 6583 &- &- & 204 & 179 & 273 & 239 & - & - & - & - \\
        {}[S~II] 6717 &- &- & $*$ & $*$ &48: & 42:  & - & - & - & - \\
        {}[S~II] 6731 &- &- & $*$ & $*$ &43: & 38:  & - & - & - & - \\
        {}[Ar V] 7005 &12: & 10: & - & - & - & -  & - & - & - & - \\
        {}[Ar III] 7135 &11: & 9: & 37 & 31 & 42 & 36  & - & - & - & - \\
        {}[Ar IV]+He~II 7175 &$*$ & $*$ & - & - & - & -  & - & - & - & - \\
        {}N I 7452 & - & - & 59 & 49 & - & -  & - & - & - & - \\
        \hline
        log({}[O~III] 5007/H$\alpha$) & - & 0.16 & - & 0.56 & - & 0.57  & - & - & - & -   \\
        log({}[O~III] 5007/H$\beta$) & - & 0.60 & - & 0.98 & - & 0.98  & - & - & - & -   \\
        log({}[N~II] 6583/H$\alpha$) & - & - & - & $-$0.17 & - &$-$0.03 & - & - & - & - \\
        log({}[S~II]/H$\alpha$) & - & - & - & - & - &$-$0.50 & - & - & - & - \\
        log({}[S~II] 6731/H$\alpha$) & - & - & - & - & - &$-$0.83 & - & - & - & - \\
        log({}[O I] 6300/H$\alpha$) & - & - & - & - & - & $-$0.94 & - & - & - & - \\
        log({}[O II] 3727/[O~III] 5007) & - & - & - & $-$0.13 & - &$-$0.80   & - & - & - & - \\
        log({}[O II] 3727/H$\beta$) & - & - & - & 0.85 & - &0.17   & - & - & - & - \\
      \hline
   \end{tabular}
   \begin{flushleft}
      $^\dag$Lines more uncertain in IDS spectra because of bright sky background. \\
      $:$Low S/N measurement.\\
      $*$Line present but too low S/N to be measured.
   \end{flushleft}
\end{table*}

\begin{figure}
   \begin{center}
      \includegraphics[scale=0.35,angle=270]{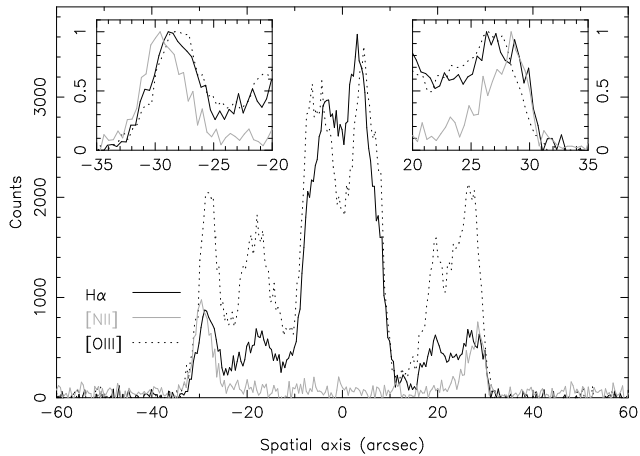}
   \end{center}
   \caption{Spatial profile of the H$\alpha$, [N~II] $\lambda$6583 and [O~III] $\lambda$5007 emission lines from the INT IDS spectrum. The insets show normalised profiles of the ends of the jets. No dereddening has been applied.}
   \label{fig:trace}
\end{figure}

The emission line ratios from separate extractions of the jet tips in Tab. \ref{tab:lines} can be compared to model predictions (Dopita 1997; Raga et al. 2008). These works aim to reproduce FLIERS as the result of shocks in the strongly photoionised medium provided by the CSPN giving the apperance of a `reverse-shock'. In ETHOS~1 the presence of [O~III] in the jet and its relatively low $T_e=$12900 K indicates strong radiative cooling is occurring in the post-shock region. We find the line ratios agree well with $\mathrm{log}\ U_2\sim-4.4$ from Dopita (1997) and the measured values of FLIERS in NGC 6543 and NGC 7009 (Raga et al. 2008). The Raga et al. (2008) models systematically underestimate the [O~III] $\lambda$5007/H$\alpha$ ratio, although this is probably due to the relatively low stellar temperatures adopted in their models. The spatial distribution of the key emission lines H$\alpha$, [N~II] $\lambda$6583 and [O~III] $\lambda$5007 are also consistent with the Raga et al. (2008) models. Figure \ref{fig:trace} presents spatial profiles of these lines where the CSPN continuum was subtracted from extractions adjacent to the lines. The results qualitatively agree with both IC 4634 and the models presented by Raga et al. (2008) with [N~II] extending $\sim$2\arcsec\ further out than the other emission lines. This effect is also evident in Fig. \ref{fig:montage}a. 

\section{Conclusions}
\label{sec:conclusion}
We have introduced ETHOS~1 (PN~G068.1$+$11.1), the first spectroscopically confirmed PN discovered from the Extremely Turquoise Halo Object Survey (ETHOS, Miszalski et al. in prep). An irradiated close binary central star was discovered with an orbital period of 0.535 days and an amplitude of 0.816 mag. The extreme amplitude is second only to Sab 41 (Miszalski et al. 2009a) and is consistent with the presence of a very hot central star that produces the highly ionised nebula ($T_e=17700$ K). The Necklace (Corradi et al. 2010) and K 1-2 (Exter et al. 2003) also share similarly large amplitudes and high-ionisation nebulae, although the absence of low-ionisation filaments in ETHOS~1 may suggest a slightly different evolutionary history. VLT FORS spectroscopy of the CSPN confirms the presence of a close binary with a large velocity separation between primary and secondary components and the nebula. The presence of N III, C III and C IV emission lines continues the trend seen in other irradiated close binary CSPN. These weak emission lines are typical of many CSPN classified as \emph{wels} in the literature and we expect many of these will turn out to be close binaries.
Further observations are required to constrain the orbital inclination, masses and radii of the binary CSPN.

A spectacular pair of jets travelling at $120\pm10$ km/s accompanies the inner nebula of ETHOS~1 adding further evidence towards the long suspected relationship between binary CSPN and jets. Their tips present emission line ratios that are consistent with shocked models of FLIERS in PNe. The fact that the jets are detached suggests a limited period of jet activity consistent with a transient accretion disk before the CE is ejected. The kinematic age of the jets ($1750\pm250$ yrs/kpc) was found to be older than the inner nebula ($900\pm100$ yrs/kpc) supporting this hypothesis. ETHOS~1 therefore continues to follow this trend previously identified in A~63 (Mitchell et al. 2007) and the Necklace (Corradi et al. 2010). Both ETHOS~1 and the Necklace have younger kinematic ages than A~63 consistent with the apparent ongoing cooling in the younger jets via [O~III] emission. A close binary engine powering jets is out of place in the literature with most models of jets incorporating orbital periods of several years (Cliffe et al. 1995; Haro-Corzo et al. 2009; Raga et al. 2009). These models may be adequate if the accretion disk is established before the companion begins its in-spiral phase as our kinematic ages suggest. Nevertheless, new models with shorter orbital periods may be necessary to model the jets in objects like ETHOS~1 which are relatively more collimated than extant model jets. 

VLT FORS imaging and MES high-resolution spectroscopy of the inner nebula of ETHOS~1 was conducted. The $X$-shaped inner nebula was reconstructed using a \textsc{shape} kinematic model consisting of bipolar outflows and a spherical nebula. The faint bipolar extensions particularly visible in the [O~III] FORS image could be an earlier ejection occurring at about the same time as the jets or alternatively may represent a bubble produced by the jets clearing out a cavity around the PN. Similar bipolar outflows are seen in A 65 (Walsh \& Walton 1996), PPA 1759$-$2834 (Miszalski et al. 2009b) and Fg 1 (Boffin \& Miszalski in prep.) suggesting they may have a binary-related origin. It may also be possible that ETHOS~1 represents a slightly more evolved state of Hb 12 (Vaytet et al. 2009) where the spherical nebula is ejected on top of a pre-existing Hb 12 system. 

\section*{Acknowledgments}
We thank the DUOPM observers supervised by Jean-Eudes Arlot and Agn\`es Acker at the OHP T120 during 20--21 July 2009 for observations that gave the first indication of $I$-band variability in ETHOS~1. 
The work of RLMC and MSG has been supported by the Spanish Ministry of Science and Innovation (MICINN) under grant AYA2007-66804. LS is in grateful receipt of a UNAM postdoctoral fellowship and is partially supported by PAPIIT-UNAM grant IN109509 (Mexico). We thank the staff at the San Pedro Mart\'ir Observatory who assisted with the observations. HB wishes to thank Yuri Beletsky and Leo Rivas for their efficient support at the VLT. We also thank the referee Orsola De Marco for her careful reading of the manuscript and useful comments. This research has made use of data obtained from the SuperCOSMOS Science Archive, prepared and hosted by the Wide Field Astronomy Unit, Institute for Astronomy, University of Edinburgh, which is funded by the UK Science and Technology Facilities Council.

\end{document}